# Chlorine-free electrochemical disinfection using graphene sponge electrodes

Giannis-Florjan Norra[a,b], Luis Baptista-Pires[a,b], Elisabeth Cuervo Lumbaque[a,b], Carles M. Borrego[a,c], Jelena Radjenovic[a,d]

[a]*Catalan Institute for Water Research (ICRA), Emili Grahit 101, 17003 Girona, Spain*

[b]*University of Girona, Girona, Spain*

[c]*Grup d'Ecologia Microbiana Molecular, Institut d'Ecologia Aquàtica, University of Girona, Girona, Spain*

[d]*Catalan Institution for Research and Advanced Studies (ICREA), Passeig Lluís Companys 23, 08010 Barcelona, Spain*

*\* Corresponding author:*

*Jelena Radjenovic, Catalan Institute for Water Research (ICRA), Emili Grahit 101, 17003 Girona, Spain*

Phone: + 34 972 18 33 80; Fax: +34 972 18 32 48; E-mail: jradjenovic@icra.cat

**Abstract**

Graphene sponge electrodes were employed for chlorine-free inactivation of *Escherichia coli* from low conductivity water. The nitrogen-doped reduced graphene oxide (NRGO) sponge anode bearing more positive charge achieved complete *E. coli* inactivation (i.e., 5 log removal) in the anode-cathode configuration at 115 A m$^{-2}$, versus 2.6 log removal using boron-doped reduced graphene oxide sponge anode. The bacteria were mainly inactivated via electrosorption and electroporation, as confirmed by the scanning electron microscopy. Storage of the electrochemically treated samples revealed further killing of the bacteria due to the damaged cell membranes. When using real tap water, 5.5 log *E. coli* removal required 5.70 kWh m$^{-3}$, which was drastically lowered to 1.38 kWh m$^{-3}$ using intermittent current and thus exploiting the capacitive properties of graphene. The developed graphene sponge anode does not form any chlorine, chlorate, or perchlorate, and holds great promise for efficient electro-disinfection without forming toxic disinfection byproducts.

# 1. Introduction

The establishment of public water disinfection and treatment is one of the greatest public health achievements of the 20$^{th}$ century. However, billions of people, mostly in rural areas, still lack access to clean drinking water and improved sanitation [1]. The United nations (UN) 2030 Agenda for Sustainable Development includes a dedicated goal on water and sanitation, with "universal and equitable access to safe and affordable drinking-water for all" as one of the targets to be reached by 2030 [2]. To achieve this goal, the often deficient and unreliable centralized drinking water treatment systems in the developing and transition countries will need to be complemented by decentralized units for water treatment and disinfection. The implementation of community or household-managed non-networked water treatment systems requires reliable and robust technologies that will ensure the microbiological quality of the treated water at low cost. In addition to this, small-scale disinfection systems need to satisfy a range of other performance criteria such as ease of use of the system, independence from the energy grid or tap pressure, environmental sustainability, socio-cultural acceptability, and potential for dissemination (e.g., low requirement for skilled personnel and maintenance in general). Among the available physico-chemical technologies that can be used in point of entry and point of use treatment devices (e.g., chemical disinfection, membrane filtration, heat and UV-based disinfection and others), no single small-scale system meets all the performance criteria needed [3,4].

Electrochemical processes offer inherent design advantages that make them very well-suited for decentralized water treatment, and may be the key technologies to ensure access to safe water, even in remote areas [5,6]. For example, electrochemical systems are modular and have a small footprint, they operate at ambient temperature and pressure, they can be solar-powered, and they do not require any addition of chemicals [5]. The production of disinfectants is achieved by water discharge at the anode (i.e., OH$^{\bullet}$, O$_3$), oxidation/reduction of the dissolved species (i.e., oxidation of chloride to HOCl/OCl$^-$, and oxygen reduction to H$_2$O$_2$), and/or anode dissolution (i.e., ferrate). Nevertheless, formation of toxic and persistent chlorinated byproducts in the presence of chloride, present in virtually all natural waters, limits the safe application of electrochemical water treatment systems [7]. Although chlorine is a very efficient disinfectant, it rapidly reacts with the organic matter to form trihalomethanes (THMs), haloacetic acids (HAAs) and other chlorinated organic byproducts [7]. At highly oxidizing anodes such as boron-doped diamond (BDD), Ti/SnO$_2$-based anodes, and Magnéli phase titanium suboxide (Ti$_4$O$_7$) electrodes, chloride is oxidized to chlorate and perchlorate, thus further compromising the quality of the treated water [8].

In our recent study [9], we demonstrated excellent electrocatalytic activity of a newly developed graphene-based sponge electrode for the oxidation of persistent organic contaminants, and at the same time, exceptionally low current efficiency for chlorine formation (i.e., only 0.04% of current efficiency in the presence of 20 mM NaCl). Moreover, graphene-based sponge electrodes had excellent electrochemical stability during both anodic and cathodic polarization, likely due to the formation of covalent C-Si and C-O bonds between graphene and SiO$_2$ [10–12], a major component of the mineral wool template employed to produce structurally stable sponge. Graphene-based sponges were produced using a simple, bottom-up, low-cost, and easily scalable method which allowed easy incorporation of atomic dopants into the structure of graphene–based coating. In this work, we evaluated the performance of boron-doped reduced graphene oxide (BRGO) and nitrogen-doped reduced graphene oxide (NRGO) coated sponges for electrochemical disinfection of low-conductivity water, including real tap water. Electrochemical inactivation of an indicator microorganism *Escherichia coli* was evaluated in one-pass, flow-through mode at different applied current densities and using different flow directions. The produced electrode materials were characterized using scanning electron microscopy (SEM), X-ray photoelectron spectroscopy (XPS), X-ray diffraction (XRD) and zeta potential analyses. To gain insight into the inactivation mechanism, we evaluated the formation of strong oxidant species (e.g., $O_3$, $H_2O_2$, $OH^•$) and performed experiments in the presence of a radical scavenger. Complete inactivation of the bacteria was verified by the Live/Dead staining experiments and additional storage experiments with the electrochemically treated samples. In addition, SEM analysis was used to characterize the bacterial morphology before and after the electrochemical treatment. Finally, experiments performed with intermittent current application revealed exceptional disinfection performance of the graphene-based sponges due to their intrinsic capacitive properties. This study opens a new route towards more energy-efficient, chlorine-free electrochemical disinfection of low-conductivity water.

## 2. Materials and methods

### 2.1 Chemicals

Graphene oxide (GO) was provided as a GO water dispersion (0.4 wt%) from Graphenea S.L. (Spain). Boric acid, urea and Live/Dead BacLight$^{TM}$ bacterial viability kits were purchased as ACS reagents from Sigma Aldrich (Spain). Mineral wool template was purchased from Diaterm (Spain). Autoclave reactor was purchased from Techinstro. Stainless steel 316 was purchased from Wire Weaving Company (the Netherlands). Chromocult® Coliform agar and the cellulose ester filters (0.45 μm) were purchased from Merck (Spain). Luria-Bertani (LB) broth and Ringer solution were purchased from Sharlab (Spain).

## 2.2 Electrode material synthesis and characterization

Graphene-based sponges were produced using a simplified bottom-up hydrothermal synthesis method reported in our previous study [9]. Briefly, pieces of mineral wool were thoroughly soaked in the specific aqueous solution of GO, namely a mix of GO (2 g L$^{-1}$) and boric acid (43 g L$^{-1}$), to produce the BRGO, and a mix of GO (2 g L$^{-1}$) and urea (300 g L$^{-1}$), for the production of NRGO, as reported elsewhere [13,14]. The soaked mineral wool was subject to hydrothermal synthesis for 12 h at 180 ºC. The resulting graphene-based sponges were then thoroughly cleaned with the Milli-Q water to remove the unbonded flakes and the remaining impurities. Zeta potential of BRGO and NRGO was measured using Zetasizer Nano ZS (Malvern Panalytical Ltd) operating with a 633 nm laser and using an aqueous solution at pH 7. A FEI Quanta FEG (pressure: 70Pa; HV: 20kV; and spot: 4) was used for the SEM analysis of the morphology of the sponges. XPS analysis was carried out in ultra-high vacuum (base pressure 1-10 mbar) using a Phoibos 150 analyzer (SPECS GmbH, Germany) with a monochromatic aluminium Kalpha x-ray source (1486.74 eV). The energy resolution of a sputtered silver foil was 0.58 eV, as determined by the FWHM of the Ag 3d5/2 peak. XRD analysis was performed with a X'pert multipurpose diffractometer using a Cu Kα radiation (l = 1.540 Å), at room temperature. It was equipped with a vertical θ–θ goniometer (240 mm radius) with fixed sample stages that do not rotate around the Ω axis as in the case of Ω–2θ diffractometers. An X'Celerator detector was utilized, which is an ultrafast X-ray detector based on real-time multiple strip technology. The diffraction pattern was recorded with a step size of 0.03 °C and a time per step of 1,000 seconds, between 4 °C and 30 ºC.

## 2.3 Electrochemical disinfection experiments

BRGO and NRGO sponges were connected to a stainless steel current collector and employed as electrodes in a flow-through, cylindrical reactor made of methacrylate (**Figure S1**). All experiments were conducted in one-pass continuous mode, with a flow rate set at 5 mL min$^{-1}$ that corresponds to surface area-normalized permeate flux of 175 L m$^{-2}$ h$^{-1}$ (LMH) and a hydraulic residence time of 3.5 min. The flow rate was controlled using a digital gear pump (Cole-Parmer). Electrochemical disinfection experiments were conducted in the chronopotentiometric mode using a multi-channel potentiostat/galvanostat VMP-300 (BioLogic, U.S.A.) and a leak-free Ag/AgCl reference electrode (Harvard Apparatus, U.S.A.). The applied anodic currents were 50, 100 and 200 mA, resulting in the anodic current densities of 29, 58 and 115 A m$^{-2}$, respectively, calculated using the projected surface area of the electrode. The employed supporting electrolyte was 10 mM phosphate buffer (Na$_2$HPO$_4$/NH$_2$PO$_4$, pH 7, electric conductivity of 1.1 mS cm$^{-1}$). Before each experiment, the reactor was first thoroughly flushed with the clean buffer solution. To investigate the impact of electrode order on the system performance, the experiments were performed using the anode-cathode (A-C) and cathode-anode (C-A) sequence at all current densities applied. To evaluate the formation of HO$^{\bullet}$, experiments were performed in 10 mM phosphate buffer, pH 7, at 200 mA of anodic current, using terephthalic acid (TA) at 20 mg L$^{-1}$ of initial concentration. TA was previously demonstrated as an ideal probe compound for electrochemically generated HO$^{\bullet}$ (second-order rate constant for the reaction of TA with HO$^{\bullet}$, k$_{TA,OH}$=4×10$^9$ M$^{-1}$ s$^{-1}$) [15], because it does not react via direct electrolysis [16] and has very low reactivity with ozone [17]. Experiments with addition of excess methanol (10 mM) were carried out at 200 mA in the NRGO(A) - NRGO(C) configuration, to determine the role of OH$^{\bullet}$ and O$_3$. The NRGO(A) - NRGO(C) configuration was also used for the following experiments. To investigate the impact of the initial concentration of bacteria, experiments were performed with 10$^5$–10$^6$ and 10$^6$–10$^7$ CFU mL$^{-1}$ of *Escherichia coli*, at 200 mA of applied anodic current. The initial concentration ranges selected were above the values expected for domestic sewage (i.e., $10^4$ CFU mL$^{-1}$) [18] to test the system performance under challenging conditions and verify whether all *E. coli* electrosorbed to the graphene sponge electrodes are also further inactivated. To exploit the capacitive properties of graphene, the reactor was operated with intermittent application of the anodic current of 200 mA using three different ON-OFF cycles: *i)* symmetrical cycles with 105 s ON - 105 s OFF equivalent to one hydraulic retention time, HRT (HRT = 3.5 min = 210 s), *ii)* symmetrical cycles of 52.5 s ON - 52.5 s OFF equivalent to half HRT (105 s), and *iii)* asymmetrical cycles with 75 s ON - 30 s OFF equivalent to half HRT (105 s). To further verify the reactor performance under realistic conditions, experiments with real tap water were performed in both continuous (100 mA) and intermittent current application modes (100 mA and 50 mA), for the cycle of 75 s ON – 30 s OFF. The ohmic drop was calculated from the ohmic internal resistance obtained in the electrochemical impedance spectroscopy (EIS). The EIS experimental data was fitted using the BioLogic EC-lab software.

**2.4 Microbiological analyses**

*Preparation of stock cultures*: An overnight culture of *E. coli* (ATCC 700078) in LB broth was used for the experiments at a concentration of ca. $10^8$–$10^9$ CFU mL$^{-1}$. Working concentrations were adjusted to approximately $10^5$–$10^6$ CFU mL$^{-1}$ by diluting the overnight culture in sterile Ringer solution for all experiments except those where the effect of the initial cell concentration was investigated. In the latter case, the initial *E. coli* concentration was adjusted to ~$10^7$ CFU mL$^{-1}$ by centrifuging the overnight culture for 10 min at 3,500 rpm (Eppendorf 5804R) and discarding the supernatant to remove chloride from the LB medium to avoid its impact on the conductivity of the supporting electrolyte. In the experiments conducted at $10^5$–$10^6$ CFU mL$^{-1}$, the initial chloride concentration was ~30 mg L$^{-1}$, derived from the LB broth. Nevertheless, analysis of chlorine formation at significantly higher chloride concentrations (i.e., 20 mM NaCl) was conducted to exclude any contribution of chlorine species in the disinfection, in accordance with our previous study [9]. The samples from electrochemical disinfection experiments were immediately preserved at 4 °C.

*Determination of cell concentration*: The abundance of *E. coli* was determined using the membrane filtration technique and serial dilutions of collected samples and Chromocult® Coliform Agar medium (Merck). Briefly, samples were serially diluted 1:10 in sterile Ringer solution followed by filtration of 1 mL of the correspondent dilutions through 0.45 µm pore size cellulose filters (Merck) that were then placed in Chromocult agar plates and incubated overnight at 37 °C. After this period, dark-blue colonies (i.e., *E. coli*) were counted and the bacterial concentration was calculated *i)* before, *ii)* immediately after treatment, and *iii)* after 16 h of storage in the dark and at 37 ºC to assess potential regrowth of electrochemically damaged *E. coli*. The regrowth experiment was conducted at 37 ºC, which is the optimal temperature for the growth of *E. coli* according to previous studies [19–21].

*Cell viability*: The viability of treated *E. coli* cells was estimated by epifluorescence microscope (Nikon, 80i) after staining with the Live/Dead bacterial viability kit (Invitrogen Molecular Probes, Inc.). 50 mL of treated water samples were centrifuged at 4,500 rpm for 20 min (Eppendorf 5804R). After removing the supernatant, they were resuspended in 5 mL of Ringer and 1 mL of resuspended sample was exposed to a mix of SYTO® 9 and Propidium Iodide (PI) mixture solution and incubated for 15 min in the dark at room temperature. Stained samples were then prepared-on slides and images were randomly captured using an image analysis software (NIS-Elements BR).

*Cell morphology*: To assess the effect of the electrochemical treatment on the morphology of *E. coli* cells, treated samples were fixed and visualized under a field emission scanning electron microscope (FESEM). Briefly, samples were fixed with a 2.5% glutaraldehyde solution buffered with 0.1M sodium cacodylate, pH 7.4, at 4ºC for 2 h. Fixed samples were then washed and progressively dehydrated with ethanol (50%, 75%, 80%, 90%, 90%, 95% and absolute ethanol × 3) at 20 minute intervals. Dehydrated samples were then dried by the critical point method (Emitech, Germany, model K 850 CPD) and finally coated with carbon in an evaporator (Emitech, Germany, model K950 Turbo Evaporator). Observations were carried out in a field emission scanning electron microscope (model S4100, HITACHI, Japan). Images were digitally recorded and processed with the Quartz PCI program (Quartz Imaging Corporation, Canada) with a resolution of 2516 × 1937 pixels.

**2.5 Chemical analysis**

The production of ozone and free chlorine that are formed at the anode was measured in C-A flow direction at 200 mA of applied anodic current, to avoid their decomposition at the cathode and enable their detection. Likewise, hydrogen peroxide that is formed at the cathode was measured in A-C flow direction, i.e., with cathode being the last electrode of the reactor, at 200 mA of applied anodic current. Ozone was determined using the N,N-diethyl-p-phenylenediamine (DPD) colorimetric method, i.e., Chlorine/Ozone/Chlorine dioxide cuvette tests LCK 310 in a chloride-free electrolyte (Hach Lange, Spain). Free chlorine formation in the presence of 20 mM NaCl was analyzed using the same test kits and represented a cumulative concentration of ozone and chlorine, because the DPD colorimetric method cannot differentiate between the contributions of individual oxidants. Hydrogen peroxide was determined using a spectrophotometric method using 0.01 M copper (II) sulphate solution and 0.1% w/v 2,9-dimethyl-1,10-phenanthroline (DMP) solution, based on the formation of Cu $(DMP)_2^+$ cation, which shows an absorption maximum at 454 nm [22]. Ozone, free chlorine, and hydrogen peroxide were measured immediately after sampling the effluent, to minimize any potential loss of these oxidants. To confirm the destruction of the bacterial cells that leads to the $K^+$ leakage, the concentration of $K^+$ ion was also measured using ion chromatography (IC) Dionex ICS-5000 HPIC system, in samples before and after treatment at 200 mA [23]. Chloride, chlorate, and perchlorate were measured by high-pressure ion chromatography (HPIC) using a Dionex ICS-5000 HPIC system. The quantification limits for the measurements of $Cl^-$, $ClO_3^-$ and $ClO_4^-$ were 0.025 mg $L^{-1}$, 0.015 mg $L^{-1}$ and 0.004 mg $L^{-1}$, respectively.

## 3. Results and discussion

### 3.1 Graphene-based sponge characterization

The optoelectronic properties of the graphene-based sponges are presented in **Figure S2**, and were similar to the materials produced in our previous study [9]. The XPS analyses revealed a C/O atomic ratio of 1.7, 3.5 and 3.6 for GO, BRGO and NRGO, respectively, demonstrating the efficiency of hydrothermal reduction of GO (**Figure S2**; **Table S1**). In the case of NRGO, 7.9% of atomic nitrogen was successfully incorporated into the RGO network (**Table S1**). The N-active sites were distributed in 0.8% pyrrolic-N (sp3 hybridized and incorporated into a five-membered ring; 399.8 eV), 4.9% pyridinic-N (sp2 hybridized and bonded to two C-atoms; 398.4 eV), 1.4% graphitic-N (401.6 eV) and 0.8% azide groups (402.7 eV) (**Table S2**). For BRGO sponge, 1.3% of atomic boron was measured. BRGO and GO contained around 1% of nitrogen originating from the commercial GO solution employed and was not a result of the reduction methodology. Due to the removal of the oxygen functional groups from the basal plane, the interlayer spacing decreased from 8.1 for GO to ≈3.5 for BRGO and NRGO (**Figure S3**). SEM analysis demonstrated a uniform coating of the mineral wool and the presence of wrinkled graphene-based sheets (**Figure S4**). The effect of atomic doping on the surface zeta-potential of graphene-based materials was also investigated at pH 7, as the surface charge can directly affect its interaction with the live bacteria. The zeta potential of the BRGO electrode was –36.2 mV, whereas a less negative value of –13.3 mV was determined for NRGO electrode (**Figure S5**), mainly due to the successful incorporation of nitrogen functional groups.

## 3.2 *E. coli* removal at graphene-based sponge electrodes: impact of atomic doping and current application

The flow direction and anode type did not have any impact on the removal of *E. coli* in the absence of current, and graphene-based sponge electrodes achieved around 0.5–0.8 log removal of *E. coli* in the initial OC run (**Figure 1A, Table S3**). The surface nanostructure was similar for both BRGO and NRGO sponge, as determined by the SEM analyses (**Figure S4**). Previous studies demonstrated antimicrobial activity for GO and RGO in suspension, with 70% and 45% of loss of viability of *E. coli*, respectively [24]. This activity was mainly assigned to the penetration of the graphene nanosheets into the cell membranes, and extraction of phospholipids [25]. Yet, "static" graphene-based coatings have a different interaction mechanism with the bacteria compared with graphene in suspension, and rely more on the oxidative stress rather than the membrane stress [26]. The exact mechanism behind the antimicrobial properties of graphene-based coatings is still under discussion. Given that both *E. coli* and graphene sponge surface were negatively charged at the experimental pH 7, electrostatic interaction between the bacteria and the electrode in the absence of current was likely minimal. In addition, *E. coli* removal was similar for BRGO-NRGO and NRGO-NRGO systems in the OC experiments, even though the determined surface zeta-potential of the BRGO electrode at pH 7 was more negative compared with the NRGO electrode (**Figure S5**). Thus, the 0.5–0.8 log removal of *E. coli* observed in the OC was likely due to the cell deposition, membrane stress caused by the direct contact with sharp nanosheets, and the ensuing reactive oxygen species (ROS)-independent oxidative stress [27]. The impact of the anode doping on the *E. coli* removal was evident only with the application of current, as explained further in the text.

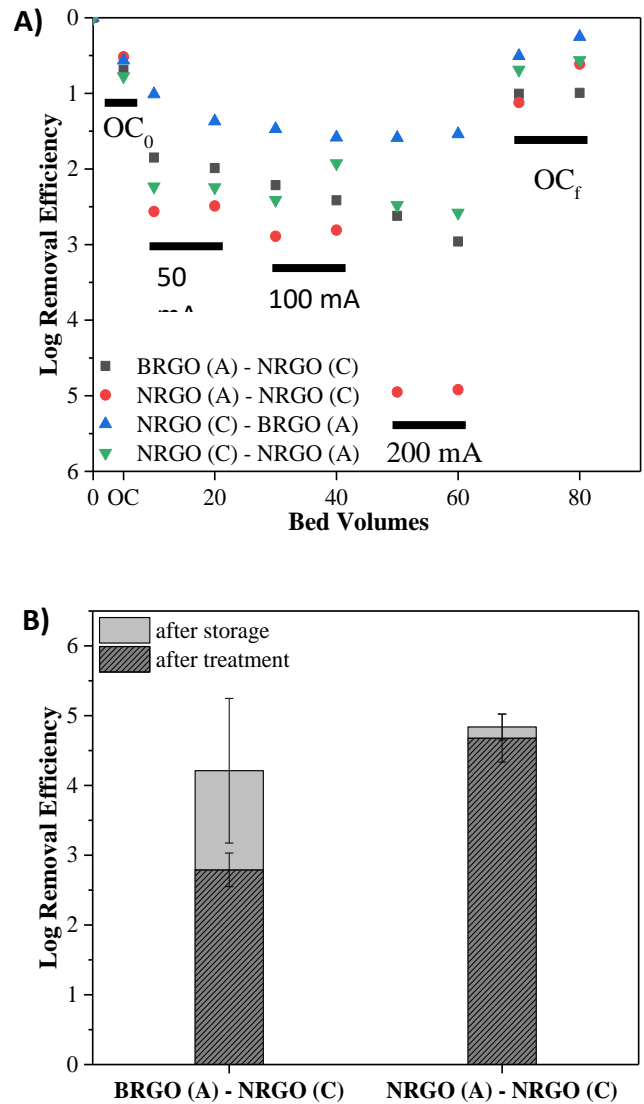

**Figure 1**. Removal of *E. coli* in phosphate buffer (10 mM, pH 7): **A)** in anode-cathode (A-C) and cathode-anode (C-A) flow directions using NRGO as cathode, and either BRGO or NRGO as anode, $OC_0$- initial open circuit run, $OC_f$- final open circuit, and **B)** in NRGO(A)-NRGO(C) system determined immediately after sampling the effluent (i.e., "after treatment"), and after storing the effluent sample at 37 °C for 16 h (i.e., "after storage"); experiment was performed using 10 mM phosphate buffer and 200 mA of anodic current.

Anodic polarization improves the removal of *E. coli* at both anodes, as can be observed from **Figure 1A**. The BRGO-NRGO system resulted in 1.9, 2.3 and 2.6 log removal in A-C flow direction at 50, 100 and 200 mA of applied anodic current, somewhat higher than the *E. coli* removal in the C-A flow direction (i.e., 1.3, 1.5, and 1.6 log removal, respectively) (**Figure 1A**). Based on the gradual return of *E. coli* effluent concentrations to the initial value once the current was switched off (i.e., final OC run, $OC_f$), the removed cells were not only electrosorbed but also inactivated. This was also supported by the storage experiments and Live/Dead analysis, as explained further in the text. Better performance of the A-C configuration using BRGO anode and NRGO cathode was also observed in our previous study focused on the removal of persistent organic contaminants [9]. This was explained by the combination of the following mechanisms: *i)* enhanced activation of $O_3$ generated at the BRGO anode by the $H_2O_2$ produced at the NRGO cathode, and *ii)* enhanced production of $H_2O_2$ via cathodic reduction of $O_2$ produced at the upstream anode [28]. The electrogeneration of $O_3$ and $H_2O_2$ was confirmed by measuring their formation at 200 mA of applied current, as explained further in the text. Lower *E. coli* removal observed at 50 and 100 mA of applied anodic current is likely a consequence of both lower electrogeneration of oxidants, and worsened interaction of the *E. coli* with the anode surface.

The *E. coli* removal was significantly improved when the dopant of the graphene-based sponge anode was changed from boron to nitrogen (**Figure 1A**). For example, at 200 mA of anodic current, *E. coli* removal in the C-A flow direction was improved from 1.6 to 2.6 log removal by substituting the BRGO anode with the NRGO anode. This was even more evident in the better performing A-C configuration, with 2.9 log removal of *E. coli* in BRGO(A)-NRGO(C) system, and up to 5 log *E. coli* removal in NRGO(A)-NRGO(C) system. Thus, the NRGO(A)-NRGO(C) configuration was capable of near complete killing of *E. coli* ($c_0=10^5-10^6$ CFU mL$^{-1}$) at 200 mA of applied anodic current (i.e., 115 A m$^{-2}$ of projected current density) from a low conductivity supporting electrolyte.

### 3.3 Inactivation mechanism of *E. coli* cells

To gain further insight into the higher disinfecting ability of the NRGO anode, we determined the formation of major oxidant species at 200 mA of applied anodic current (**Table S4**). The amount of ozone formed is presented for the C-A configurations as this configuration avoids its loss at the cathode and enables its detection. The anodes likely had the same electrocatalytic activity towards ozone formation regardless of the flow direction because electrochemical ozone generation is based on water electrolysis ($3H_2O \rightarrow O_3 + 6H^+ + 6e^-$, $E^0=1.51$ V). The NRGO and BRGO anode formed around 0.12 mg L$^{-1}$ and 0.14 mg L$^{-1}$ of ozone, respectively, at 200 mA. Although the amounts of ozone at graphene sponge anodes are well below the typical disinfection doses applied in drinking water treatment (i.e., 1.5 – 3 mg L$^{-1}$) [29], ozone has high reactivity towards *E. coli* and thus it contributed towards its inactivation [30,31]. Similar electrocatalytic activity of the BRGO and NRGO anode towards ozone formation suggests that other mechanisms were determining for the 2-log higher inactivation of *E. coli* using NRGO isntead of the BRGO anode (**Figure 1A**). In addition, the inertness of the graphene sponge anodes towards chlorine generation was confirmed by performing electrolysis in the presence of 20 mM NaCl, which revealed very low free chlorine concentration at 200 mA of anodic current (115 A m$^{-2}$) for both BRGO (0.41 mg L$^{-1}$) and NRGO anode (0.03 mg L$^{-1}$). Also, we did not measure any chlorate and perchlorate formation in these conditions, in accordance with our previous study [9].

Besides ozone, H$_2$O$_2$ was measured in the A-C configurations in which oxygen produced at the upstream anode enhances the cathodic formation of H$_2$O$_2$ at the downstream cathode ($O_2 + 2H^+ + 2e^- \rightarrow H_2O_2$, $E°= -0.67$ V) [9]. In accordance with our previous study [9], low concentrations of $H_2O_2$ were measured in both BRGO(A)-NRGO(C) and NRGO(A)-NRGO(C) systems (i.e., 0.88 and 0.66 mg L$^{-1}$ at 200 mA, **Table S5**). These amounts likely represented residual and not formed $H_2O_2$ due to its decomposition to OH$^\bullet$ at the N-active sites of the NRGO cathode [32], as well as its reaction with the anodically generated ozone ($2O_3 + H_2O_2 \rightarrow 2OH^\bullet + 3O_2$). Inactivation of *E. coli* with $H_2O_2$ is generally inefficient, as significantly higher peroxide concentrations (i.e., 34 mg L$^{-1}$) did not inactivate *E. coli* [33]. On the other hand, OH$^\bullet$ are capable of efficient *E. coli* inactivation [33]. In the experiments with TA, the estimated steady state concentration of OH$^\bullet$ was the same for both BRGO(A)-NRGO(C) and NRGO(A)-NRGO(C) systems (i.e., $1.5\pm0.09\times10^{-13}$ M, **Table S5**). The role of OH$^\bullet$ in the inactivation of *E. coli* in graphene-based sponge electrodes was investigated by adding excess methanol. However, the removal of *E. coli* was somewhat higher (i.e., 5.3–6.0 log removal) compared with the removal in the absence of methanol (5.1 log removal) (**Figure 2, Table S6**). Control experiments showed that the selected concentration of methanol (i.e., 10 mM) was not toxic to *E. coli*, which is in accordance with the literature [34]. Nevertheless, the toxicity of methanol to *E. coli* may be increased in the presence of the electric field. Recently, Xie et al. [35] reported greater intracellular diffusion of ozone after the application of electric field, which was explained by the formation of pores and disruption of the integrity of the cell membrane in the electric field. Herraiz-Carboné et al. reported recently that the electrogenerated ozone can attack the genetic material inside the bacterial cells, thus inactivating the *Klebsiella pneumoniae* [36]. Once the electric field facilitates the penetration of methanol in the bacterial cell, it can accumulate within the cell and severely affect its function, causing cell death [37]. This can explain a slightly higher removal of *E. coli* with the addition of a radical scavenger, but at the same time, it makes it difficult to determine the contribution of OH$^\bullet$ to *E. coli* inactivation.

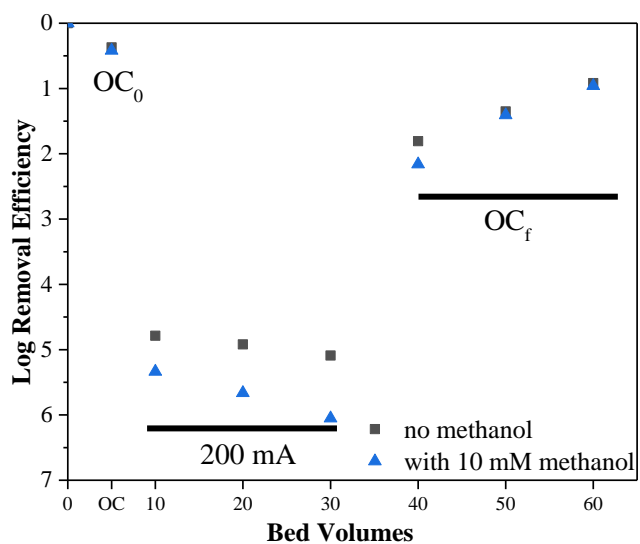

**Figure 2**. Removal of *E. coli* in 10 mM phosphate buffer (square) and with the addition of 10 mM methanol (triangle), in the initial $OC_0$, 200 mA of applied anodic current, and OC applied at the end of experiment ($OC_f$).

Given the similar electrocatalytic activity of the BRGO and NRGO anodes towards ozone generation (**Table S4**), and similar amounts of $H_2O_2$ formed at the NRGO in the two investigated configurations (**Table S5**), higher removal of *E. coli* when using NRGO anode can be explained by its enhanced electrosorption at the NRGO anode surface, resulting in its more complete inactivation than in the case of the BRGO anode. Electrosorption of the negatively charged bacterial cells at the positively charged anode surface and their subsequent inactivation due to electroporation is an important mechanism of cell inactivation at carbon-based electrodes; electroporation can lead to loss of the structural integrity of the cell membrane and thus cell death [38]. Previous study identified oxidative stress as the predominant antibacterial mechanism of N-doped reduced graphene oxide in suspension [39]. Pyridinic-N and pyrrolic-N identified in the NRGO sponge (4.9% and 0.8%, respectively, section 3.1) are located on the graphene edges and provide a p-type doping, i.e., electron-deficient sites, whereas graphitic-N (1.4%) is an n-type dopant representing an electron-rich site. In the case of BRGO, boron atom is also a p-type dopant, yet the percentage of boron incorporated into graphene is significantly lower (1.3%) compared with the pyridinic-N in NRGO (4.9%). Thus, the NRGO anode attracts more the negatively charged *E. coli* due to more electron-deficient sites (i.e., 5.7% of pyridinic and pyrrolic-N in NRGO anode, versus 1.3% of boron in BRGO anode) and thus higher localized positive charge accumulated at the NRGO sponge surface during anodic polarization.

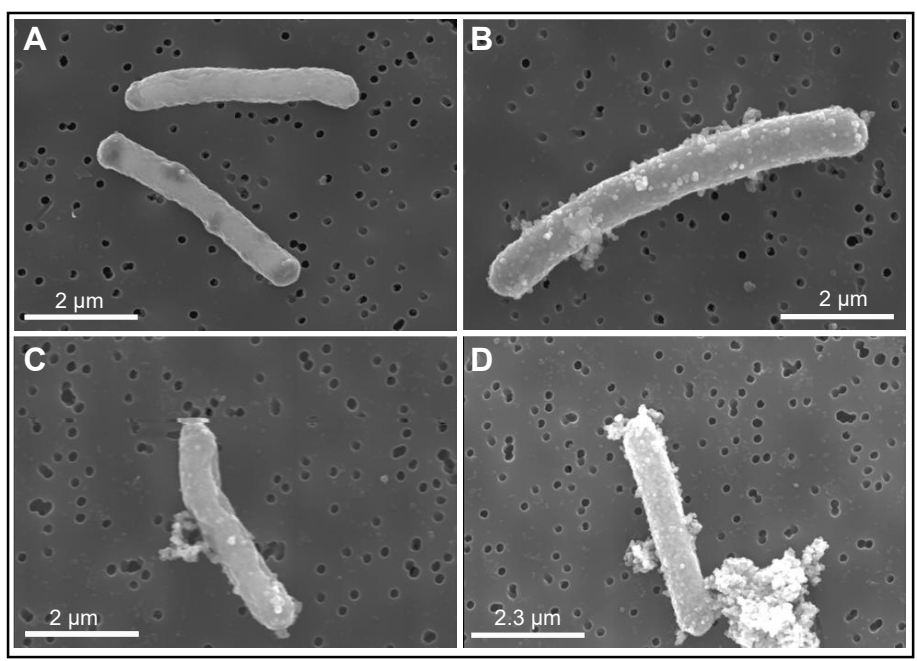

**Figure 3.** SEM images of *E. coli* before (**A**) and after (**B-D**) treatment

FESEM images of the electrochemically treated *E. coli* cells show the damage induced in the cell membranes *via* electroporation (**Figure 3**). After electrochemical treatment, many cells show protrusions of intracellular content through pores (**Figure 3B-D**) and cell walls show sunken surface (**Figure 3C**). The formation of pores in the cell membrane is a typical outcome of electroporation that results in bacterial inactivation [40]. Similar low-voltage electroporation was previously reported for carbon nanotube (CNT)-based sponges [41]. Furthermore, the analyses of potassium showed an increase in K$^+$ concentration, from $0.58 \pm 0.02$ mg L$^{-1}$ in the initial solution to $1.18 \pm 0.29$ mg L$^{-1}$ after treatment (**Figure S6**). The K$^+$ ion has multiple roles in the *E. coli* cells, serving as an essential osmotic solute, activator of intracellular enzymes, regulator of intracellular pH, and a second messenger to stimulate accumulation of compatible solutes [42]. Leakage of $K^+$ results from either the change in cell permeability, or complete destruction of the cell, and was previously reported in electrochemical disinfection using BDD anode in sulfate-based solution [23].

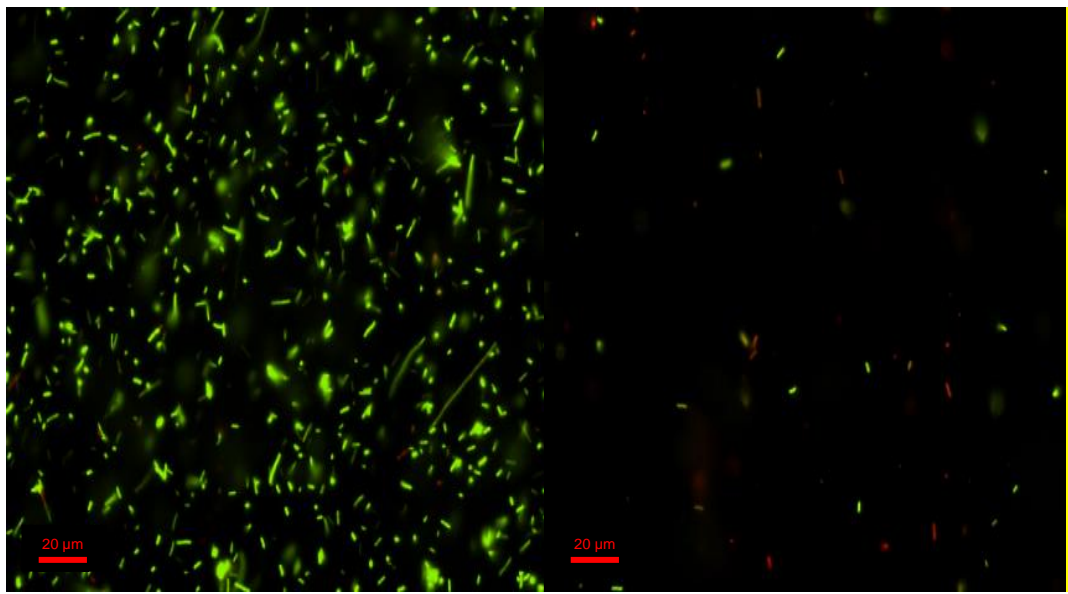

**Figure 4.** Micrograph of *E. coli* cells before (left) and after (right) treatment using the Live/Dead bacterial viability kit. Green- and red-stained cells correspond to cells with intact and damaged membranes, respectively.

The impairment of membrane integrity of *E. coli* cells was confirmed with the Live/Dead analyses of the electrochemically treated sample, with a drastic decrease in the number of viable cells (green-stained) and the appearance of damaged ones (red-stained) in the sample treated at 200 mA (**Figure 4**). Remarkably, the low number of red-stained cells in the sample corresponding to 5 log *E. coli* removal are likely due to complete cell lysis for most inactivated cells, making staining with PI unfeasible. Furthermore, samples treated at 200 mA in the BRGO(A)-NRGO(C) and NRGO(A)-NRGO(C) systems showed further decrease in *E. coli* after storage experiments conducted at optimum growth temperature (37 ºC) for 16 h (**Figure 1B**). This decrease was more obvious for the BRGO(A)-NRGO(C) system due to the presence of more *E. coli* cells that withstand the treatment under this configuration (i.e., ~$10^2$ CFU mL$^{-1}$). Further cell death with storage time demonstrates irreparable damage of cell walls after the electrochemical treatment. Near complete removal of *E. coli* in the NRGO(A)-NRGO(C) system made it difficult to see further decrease in *E. coli* concentration. Yet, it confirmed that there was no regrowth of the bacteria during storage.

**3.4 Impact of the initial *E. coli* concentration on disinfection performance**

When the initial concentration of *E. coli* was increased to $10^6$–$10^7$ CFU mL$^{-1}$, the resulting removal of *E. coli* increased from 5.1 ± 0.3 log removal to 5.8 ± 1.2 log removal (**Figure S7**). Thus, one-pass operation of the graphene sponge-based electrochemical system at 200 mA and with the HRT of 3.5 min was still sufficient to achieve near complete disinfection even at higher initial *E. coli* concentrations, and without any contribution by active chlorine species, which is the common disinfectant employed in electrochemical systems.

**3.5 *E. coli* removal with intermittent current application**

In the scientific literature, one of the main applications of the graphene-based materials has been in the energy storage field, as supercapacitors, due to their pseudocapacitive behavior, high specific surface area, tunable porosity and hierarchical arrangement, in addition to low cost and ease of synthesis [43,44]. Given the ability of these materials to perform continuous and fast charge/discharge cycles, we investigated the disinfection capability of the best-performing configuration, NRGO(A)-NRGO(C), under intermittent current application. As illustrated in **Figure 5A**, the obtained removal of *E. coli* was largely dependent on the duration of the ON and OFF cycles. When the current pulse was symmetrical and total cycle time equivalent to one HRT (i.e., ON and OFF cycle duration of half HRT, 105 s each), only up to 3.3 log removal of *E. coli* was achieved, compared with up to 5.1 log removal under continuous current.

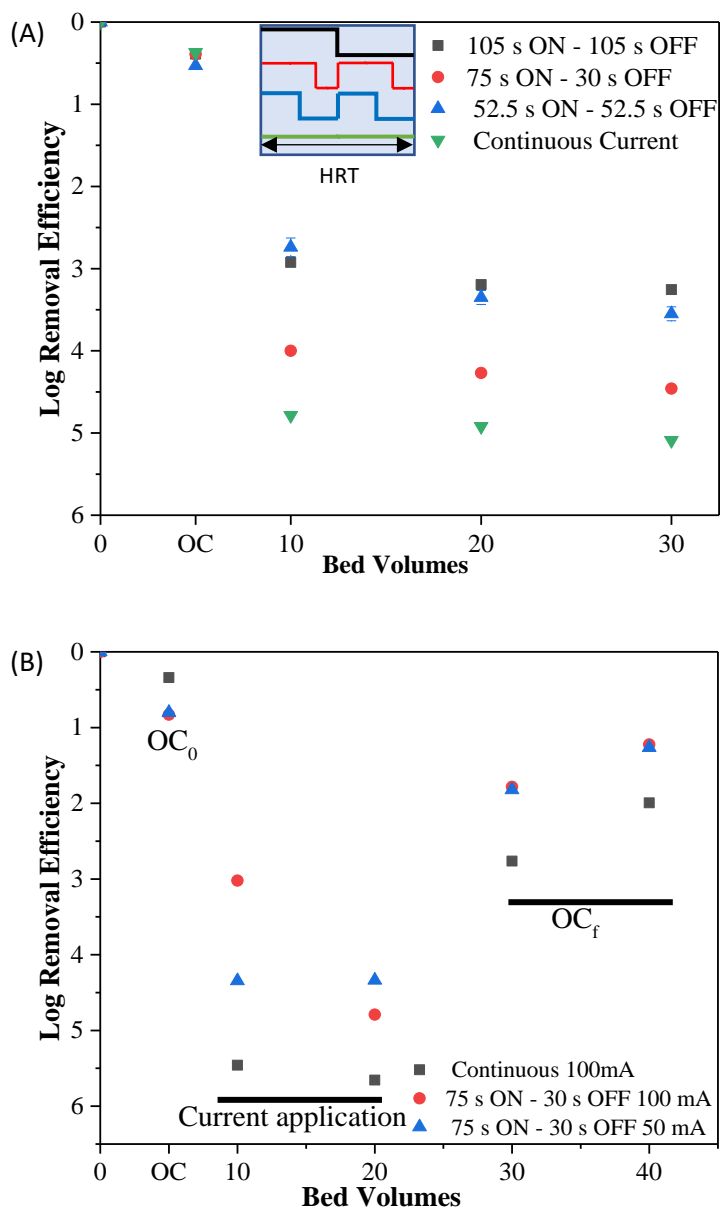

**Figure 5.** Removal of *E. coli* in **A)** phosphate buffer (10 mM, pH 7) at continuous current and with intermittent current application, both at 200 mA, with varying durations of ON and OFF cycles, and **B)** tap water at continuous current (100 mA) and with intermittent current application at 50 mA and 100 mA.

Given that graphene-based materials undergo fast charging but also fast discharging [43,44], it is likely that the duration of the OFF cycle was too long for the graphene-based sponges to maintain the sufficient current density and contribute to disinfection during the OFF stage. Very similar results and up to 3.6 log removal of *E. coli* was obtained when the reactor was operated using two symmetrical pulses within one HRT (i.e., ON and OFF cycles of 52.5 s each), and in this case may have been a consequence of too short ON cycles that did not allow sufficient electrosorption of *E. coli* onto the graphene sponge electrodes, thus leading to their lower inactivation (**Figure 5A, Table S7**). Thus, we applied asymmetrical pulses with total duration of one HRT, and 75 s of ON cycle/30 s of OFF cycle. This operation resulted in up to 4.5 log removal of *E. coli*. This result is similar to that recorded with continuous current (i.e., up to 5.1 log removal) but the application of intermittent current enabled a significantly lower energy consumption (i.e., 4.82 kWh m$^{-3}$) compared with the continuous current (i.e., 8.83 kWh m$^{-3}$) (**Table S8**). Thus, pseudocapacitive behavior of graphene-based sponge electrodes may allow saving energy by applying intermittent current and without compromising the disinfection performance.

### 3.6 *E. coli* removal from real tap water

The performance of the NRGO(A)-NRGO(C) system was investigated with real tap water (characteristics in **Table S9**) spiked with *E. coli* ($10^6$–$10^7$ CFU mL$^{-1}$). Electrochemical disinfection of tap water was performed using continuous current application at 100 mA, resulting in the ohmic-drop corrected anode potential of 2.4 V/SHE (recorded anode potential of 5.8 V/SHE), same as in the case of the more conductive 10 mM phosphate buffer at 200 mA (i.e., ohmic-drop corrected potential of 2.4 V/SHE, recorded potential of 5.4 V/SHE) (**Table S10**). The obtained *E. coli* removal was similar for the two experiments, with up to 5.5 log removal obtained for tap water (**Figure 5B**). Thus, phosphate anions did not impact negatively the electrosorption and electrochemical inactivation of *E. coli* at the graphene-based sponge electrodes. Application of 100 mA of anodic current in 75 s ON – 30 s OFF mode resulted in up to 4.9 log removal of *E. coli* (**Figure 5**). However, although the experiment in tap water using continuous application of 100 mA could be performed without any difficulties, experiment performed in the intermittent mode rapidly led to a large increase in both anode and cathode potentials, resulting in the total cell potential >34 V (limit of the potentiostat unit employed) and frequent process failure in the replicate experiments (**Table S10**). Limited ion diffusion in the low conductivity tap water, as well as larger ohmic losses will drive the electrode potentials upwards. The capacitance and rate capability of the graphene-based electrodes will be greatly affected by the diffusion of inserted and de-inserted ions, with aqueous electrolytes with lower ionic conductivity and mobility having slower charging and discharging rates [45]. Thus, application of the intermittent current may result in the overall higher total cell potential compared with the continuous current due to the slower discharge during the short 30 s OFF stages of the cycle [46,47]. By lowering the current to 50 mA, operation of the reactor in 75 s ON – 30 s OFF cycles was more stable and up to 4.5 log removal of *E. coli* was achieved (**Figure 5**). Electrochemical disinfection of real tap water using the intermittent current mode led to significant savings in the energy consumption, which was decreased from 5.70 kWh m$^{-3}$ in continuous current mode (up to 5.5. log removal) at 100 mA, to 4.55 kWh m$^{-3}$ in the intermittent current mode at 100 mA (4.9 log removal), and 1.38 kWh m$^{-3}$ in the intermittent current mode at 50 mA of anodic current (4.5 log removal). These results compare favorably to the energy consumption of electrochemical disinfection systems reported in literature, even including chlorine-mediated disinfection, and yet are achieved in a chloride-free electrolyte. For example, electro-disinfection of the liquid fraction of blackwater required around 4.4 kWh m$^{-3}$, although the process was chlorine-mediated, and the influent conductivity was significantly higher compared with the tap water used in our study [48]. Electrochemical disinfection of *E. coli* in 30 mM Na$_2$SO$_4$ and platinum-based anodes achieved 4 log *E. coli* removal and required 6.3 kWh m$^{-3}$ [49]. BDD anodes can achieve efficient electrochemical disinfection of wastewater at low current densities (e.g., 1-50 A

m$^{-2}$) and thus low energy consumptions (e.g., 0.1-1.1 kWh m$^{-3}$) [50,51], but the process relies on the generation of chloramines, which in turn may lead to the formation of N-nitrosodimethylamine (NDMA), a highly toxic and carcinogenic disinfection byproduct [52]. Also, electrochemical disinfection of sewage effluent using an integrated electrodisinfection-electrocoagulation system equipped with a BDD anode and sacrificial iron electrodes required 2.24 kWh m$^{-3}$ for complete inactivation of *E. coli*, whereas the formation of toxic chlorinated byproducts was limited by applying low current densities (<7 A m$^{-2}$) and electrical charges (<0.07 kAhm$^{-3}$) [53]. In our study, although the employed tap water contained low concentration of chloride (20 mg L$^{-1}$, **Table S9**), it was not oxidized at the anode to chlorine or other chlorinated species (chlorate, perchlorate), as was previously demonstrated when using graphene sponge anodes and significantly higher chloride concentrations [9]. Given that the graphene sponge electrodes do not oxidize chloride, the formation of disinfection byproducts such as THMs and HAAs typical of chlorine-based disinfection is excluded and makes the electrodes more versatile for use with different water matrices and current densities.

## 4. Conclusions

Electrochemical disinfection was investigated in a one-pass flow-through electrochemical system equipped with graphene sponge electrodes, using low conductivity supporting electrolyte and *E. coli* as an indicator microorganism. The main conclusions of the current study are the following:

- The order of the electrodes had a determining impact on the reactor performance, with anode-cathode flow direction yielding on average two-fold higher *E. coli* log removal compared with the cathode-anode configuration.

- Atomic doping of the graphene sponge was crucial to further enhance the disinfection performance. The NRGO sponge anode bearing more positive charge achieved complete inactivation of *E. coli* (i.e., 5 log removal) in a chloride-free electrolyte, versus 2.6 log removal of *E. coli* using BRGO sponge anode, both at the projected anodic current density of 115 A m$^{-2}$.

- The main inactivation mechanism of *E. coli* was based on the electrosorption of bacteria to the anode surface and cell lysis due to the disruption of the cell walls by electroporation. The changes in the cell morphology were confirmed by the FESEM analyses and Live/Dead staining experiments. Storage experiments conducted with the electrochemically treated samples also confirmed the damaged cell walls and no regrowth of *E. coli* after treatment. The electrogenerated ozone and hydroxyl radicals were measured, and these species likely contributed to *E. coli* inactivation.

- Electrochemical system equipped with the NRGO anode and cathode was employed for tap water disinfection at 58 A m$^{-2}$, and achieved up to 5.5 log removal of *E. coli* with an energy consumption of 5.70 kWh m$^{-3}$. This energy consumption could be further lowered to 1.38 kWh m$^{-3}$ by applying current in an intermittent mode (i.e., in ON/OFF pulses of 29 A m$^{-2}$), without significant decrease in the inactivation efficiency (i.e., 4.5 log removal), due to the intrinsic capacitive properties of graphene-based materials.

Given the low electrocatalytic activity of the graphene sponge anode towards chloride oxidation (current efficiency <0.1%), and the fact that no chlorine, chlorate, and perchlorate was detected in the system even at high current densities in the presence of high chloride concentration (i.e., 20 mM NaCl), the present study demonstrates great potential of the developed materials for water disinfection without forming disinfection by-products typically observed in chlorine-based disinfection. The estimated cost of graphene-based sponge is only €23 per m² of projected surface area, given the lower concentration of the GO dispersion used in the synthesis compared with our previous study [9]. This is extremely competitive compared with the state-of-the-art BDD anodes (~€4,200 per m²) and dimensionally stable anodes (DSAs) (€3,000 per m²) [54]. Given that they are produced using a simple and scalable production method, graphene-based sponge electrodes hold great promise for practical applications of electrochemical water treatment systems and their deployment in the developing countries and rural communities. For example, a 1 m² solar panel can produce up to 208 kWh/year (i.e., 569 Wh per day) of energy, assuming 16% efficiency [55]. For the treatment of tap water, the system developed in this study and connected to a 1 m² solar panel would produce 118 L per day of disinfected water, which could be further increased to 412 L per day using intermittent current supply. Thus, solar-powered electrochemical disinfection using low-cost graphene sponge electrodes may provide clean water for impoverished communities not connected to the energy grid.

## Acknowledgments

The authors would like to acknowledge ERC Starting Grant project ELECTRON4WATER (Three-dimensional nanoelectrochemical systems based on low-cost reduced graphene oxide: the next generation of water treatment systems), project number 714177. ICRA researchers thank funding from CERCA program.

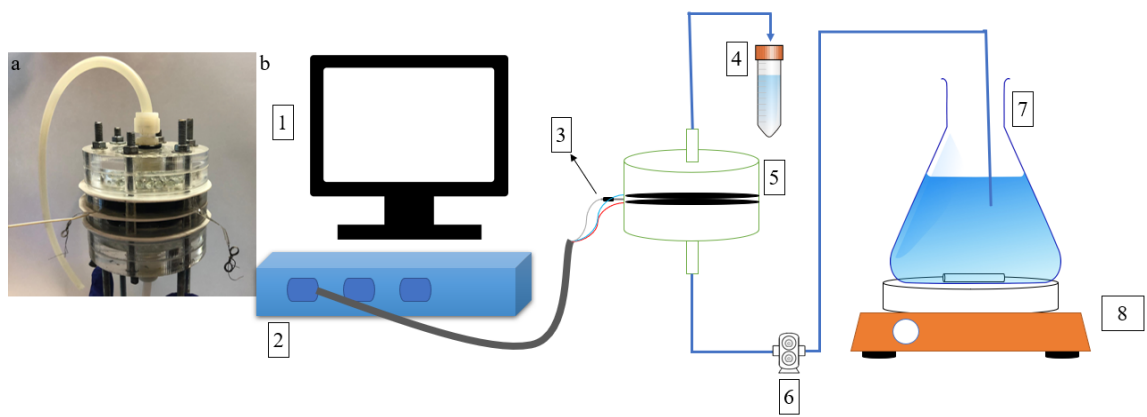

**Figure S1. a)** Photo of the flow-through reactor, and **b)** Scheme of the experimental set-up used (1) computer for the control of current and data acquisition, (2) potentiostat, (3) reference electrode, (4) sample collector, (5) electrochemical reactor, (6) digital gear pump, (7) influent reservoir, and (8) magnetic stirrer.

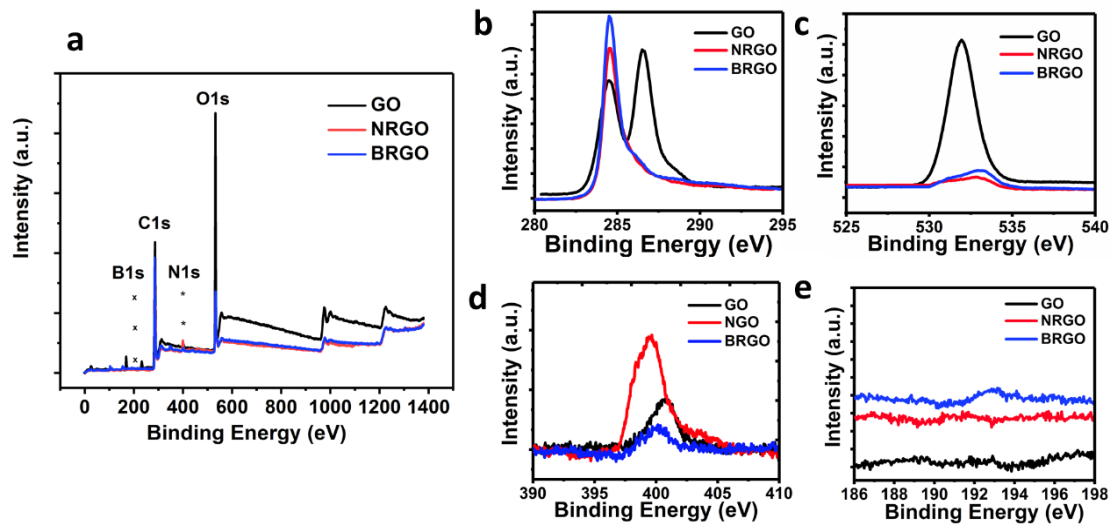

**Figure S2**. X-ray photoelectron spectroscopy (XPS) analysis of the graphene-based sponges with reduced graphene oxide (RGO) coating: **a)** wide region, **b)** C1s, **c)** O1s, **d)** N1s and **e)** B1s XPS spectra of GO, BRGO and NRGO.

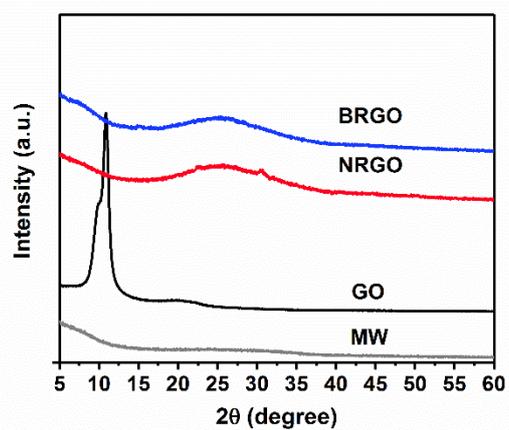

**Figure S3**. X-ray diffraction (XRD) patterns of the mineral wool, graphene oxide (GO) employed in the synthesis, and the BRGO and NRGO sponges.

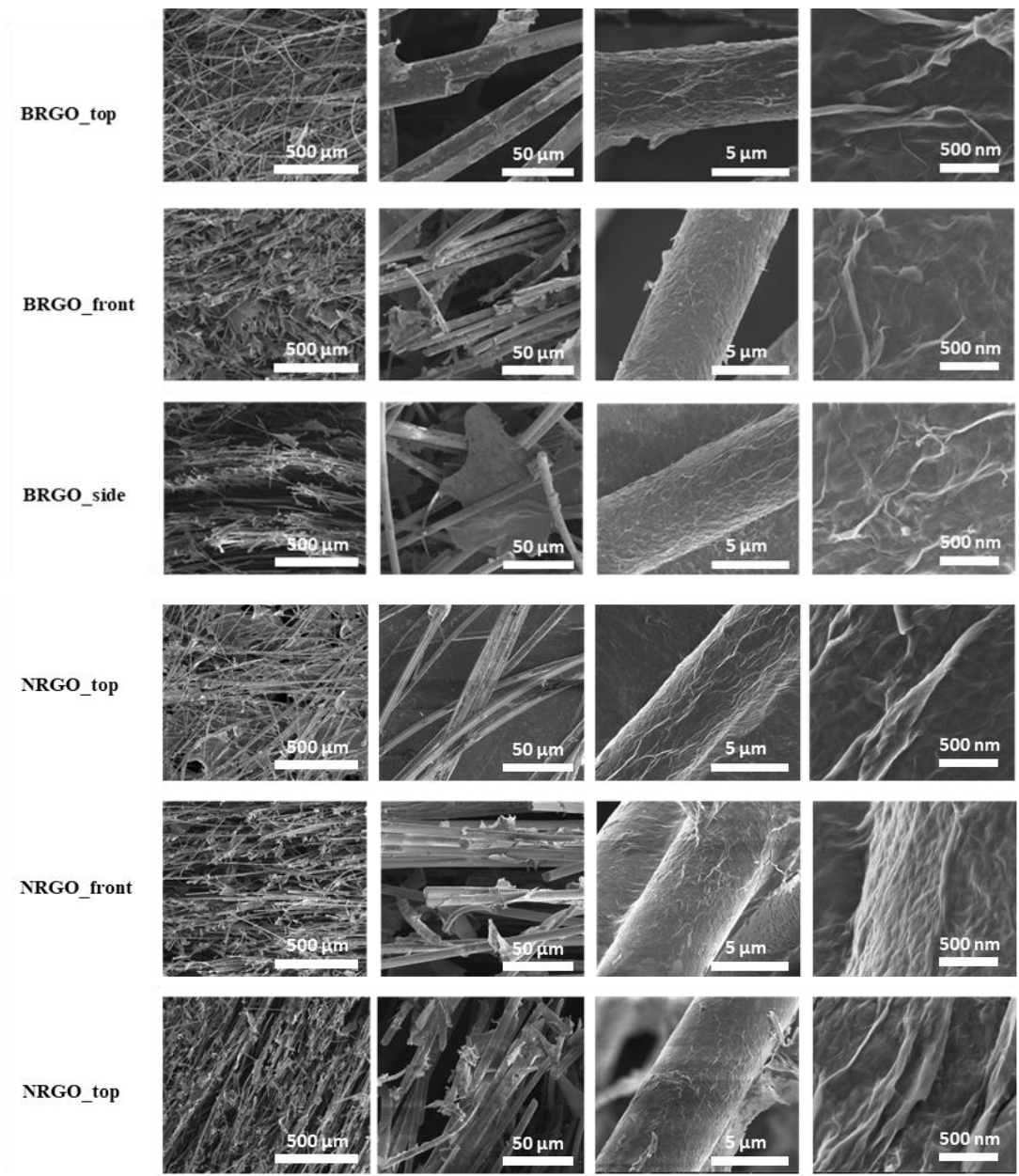

**Figure S4**. SEM images of BRGO and NRGO sponges.

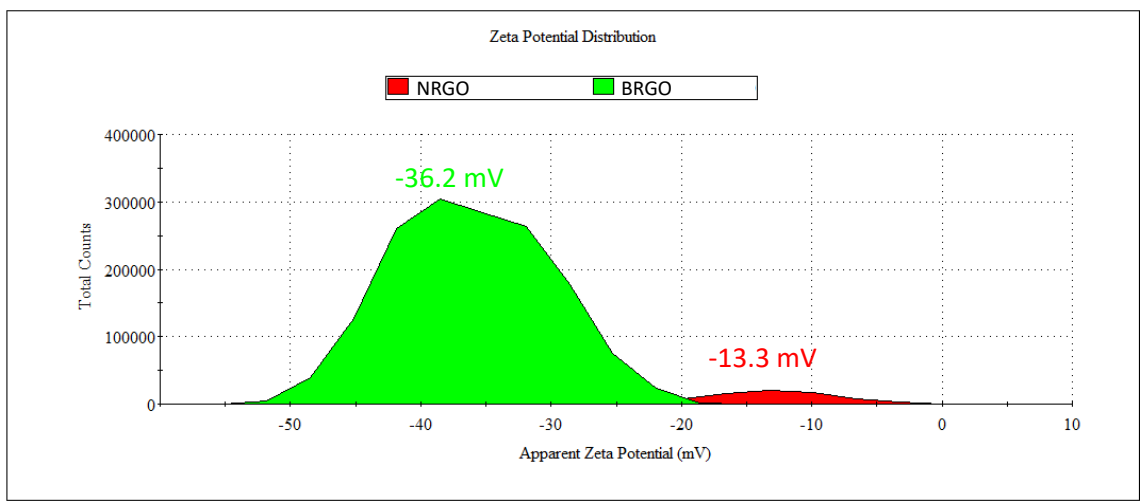

**Figure S5**: Zeta potential distribution of the BRGO and NRGO sponges.

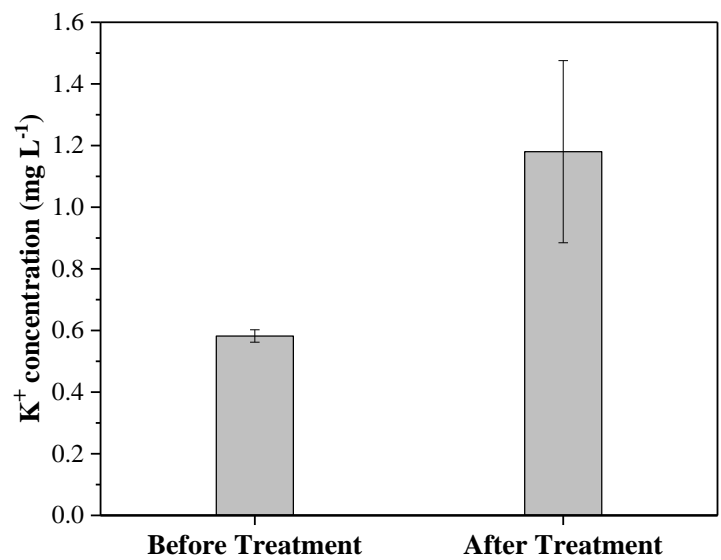

**Figure S6**. $K^+$ concentration in the 10 mM phosphate buffer amended with $10^7$ CFU mL$^{-1}$ of *E. coli* before and after electrochemical treatment at 200 mA with the NRGO (A) - NRGO (C) configuration.

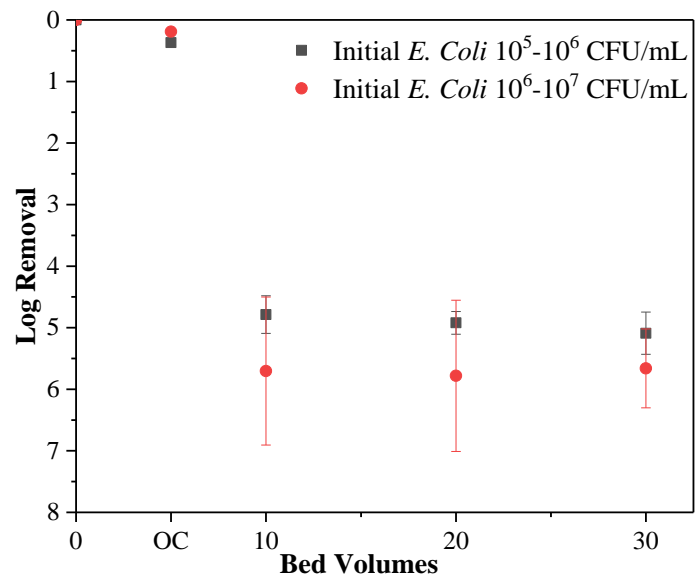

**Figure S7**. Removal of *E. coli* in phosphate buffer (10 mM, pH 7) at the initial concentration of *E. coli* of $10^5$-$10^6$ (square) and $10^7$ (circle) CFU mL$^{-1}$.

**Table S1.** The atomic content of the GO precursor solution and synthesized graphene-based sponges, as determined by the XPS analyses.

|       | GO   | NRGO | BRGO |
|-------|------|------|------|
| C (%) | 62.6 | 72.1 | 75.6 |
| O (%) | 36.9 | 19.9 | 21.8 |
| N (%) | 0.9  | 7.9  | 1.2  |
| B (%) | 0    | 0    | 1.3  |

**Table S2.** Percentage of functional groups of C1s, O1s, and N1s in the XPS spectra of the GO precursor solution and synthetized graphene-based sponges, as determined by the deconvolution of peaks in the XPS analyses.

|  | C1s (%) | | | | O1s (%) | | | N1s (%) | | | |
| --- | --- | --- | --- | --- | --- | --- | --- | --- | --- | --- | --- |
|  | *C-C 284.5eV* | *C-N 285.6eV* | *C-O 286.9eV* | *COOH 288.5eV* | *C-O 531eV* | *C=O/ C-N-O 533eV* | *536.3eV* | *Pyridinic N 398.4eV* | *Pyrrolic N 399.8eV* | *Graphitic N 401.6eV* | *Azide 402.7eV* |
| **GO** | 39.1 | 4.9 | 49.7 | 6.35 | 96.4 | 3.6 | 0 | 0 | 100 | 0 | 0 |
| **NRGO** | 67.6 | 18.6 | 8.61 | 5.05 | 23.4 | 76.6 | 0 | 9.6 | 61.8 | 18.3 | 10.2 |
| **BRGO** | 76.4 | 17.4 | 4.5 | 1.7 | 36.6 | 63.4 | 0 | 21.4 | 43.1 | 22.8 | 12.8 |

**Table S3.** Log removal efficiency (mean ± standard deviation) of *E. coli* in varying reactor configurations and flow directions, at the samples bed volumes (BV). $OC_0$-open circuit conducted at the beginning of each experiment. $OC_f$-final open circuit, conducted at the end of each experiment (i.e., after the application of current).

| Current | BRGO (A) - NRGO (C) | NRGO (A) - NRGO (C) | NRGO (C) - BRGO (A) | NRGO (C) - NRGO (A) |
|---|---|---|---|---|
| $OC_0$ | 0.69±0.43 | 0.52±0.11 | 0.56±0.19 | 0.78±0.02 |
| 50 mA, 10 BV | 1.85±0.76 | 2.56±0.61 | 1.01±0.10 | 2.23±0.06 |
| 50 mA, 20 BV | 1.99±0.76 | 2.49±0.45 | 1.37±0.53 | 2.24±0.06 |
| 100 mA, 30 BV | 2.21±0.51 | 3.11±0.72 | 1.47±0.46 | 2.41±0.07 |
| 100 mA, 40 BV | 2.42±0.53 | 2.81±0.55 | 1.58±0.87 | 1.93±1.2 |
| 200 mA, 50 BV | 2.62±0.07 | 4.54±0.16 | 1.59±0.52 | 2.48±1.16 |
| 200 mA, 60 BV | 2.96±0.73 | 4.92±0.18 | 1.54±0.43 | 2.58±1.38 |
| $OC_f$, 70 BV | 1.00±0.56 | 1.12±0.65 | 0.50±0.24 | 0.69±0.55 |
| $OC_f$, 80 BV | 0.99±0.57 | 0.61±0.68 | 0.25±0.12 | 0.56±0.37 |

**Table S4.** Concentrations of ozone and free chlorine measured using BRGO and NRGO anode in C-A configuration using phosphate buffer (10 mM, pH 7), and in the presence of 20 mM NaCl, respectively, at 200 mA.

|  | $O_3$ (mg L$^{-1}$) | Free Chlorine (mg L$^{-1}$) |
|---|---|---|
| NRGO (C) – NRGO (A) | 0.12 ± 0.01 | 0.06 ± 0.003 |
| NRGO (C) – BRGO (A) | 0.14 ± 0.006 | 0.41 ± 0.19 |

**Table S5.** Concentration of $H_2O_2$ and quasi steady-state concentration of OH$^\bullet$ ([OH$^\bullet$]$_{ss}$)* measured using BRGO and NRGO anode in A-C configuration using phosphate buffer (10 mM, pH 7) at 200 mA

|  | $H_2O_2$ (mg/L) | [OH$^\bullet$]$_{ss}$ (M) |
|---|---|---|
| NRGO (A) – NRGO (C) | 0.66±0.03 | 1.54±0.09×10$^{-13}$ |
| BRGO (A) - NRGO (C) | 0.88±0.22 | 1.48±0.06×10$^{-13}$ |

*$[OH^\bullet]_{ss} = \frac{k_{TA}}{k_{TA,OH}}$ where $k_{TA,OH}$ (4×10$^9$ M$^{-1}$ s$^{-1}$) [1] is the second-order rate constant for OH$^\bullet$ with terephthalic acid (TA), and $k_{TA}$ (s$^{-1}$) is the pseudo-first rate constant of TA decay [2,3].

**Table S6.** Log removal efficiency (mean ± standard deviation) of *E. coli* in the continuous current experiment at 200 mA (10 mM phosphate buffer, pH 7), with and without the addition of 10 mM methanol.

| Bed Volumes | No added methanol | With 10 mM methanol |
|---|---|---|
| OC$_0$, 10 BV | 0.37±0.11 | 0.42±0.23 |
| 10, X BV | 4.79±0.60 | 5.34±0.91 |
| 20, X BV | 4.92±0.18 | 5.66±0.44 |
| 30, X BV | 5.09±1.23 | 6.05±0.11 |
| 40, X BV | 1.81±0.23 | 2.16±0.13 |
| 50 OC$_f$, X BV | 1.35±0.92 | 1.41±0.02 |
| 60 OC$_f$, X BV | 0.92±0.27 | 0.96±0.06 |

**Table S7.** Log removal efficiency (mean ± standard deviation) of *E. coli* in the continuous and intermittent current experiments in 10 mM phosphate buffer using the NRGO(A)–NRGO(C) configuration at 200 mA.

| Bed Volumes | Continuous | 105 s ON 105 s OFF | 75 s ON 30 s OFF | 52.5 s ON 52.5 s OFF |
|---|---|---|---|---|
| OC$_0$ | 0.37±0.03 | 0.39±0.03 | 0.39±0.03 | 0.53±0.06 |
| 10 | 4.79±0.31 | 2.93±0.54 | 4.00±0.04 | 2.74±0.11 |
| 20 | 4.92±0.18 | 3.20±0.11 | 4.27±0.20 | 3.35±0.08 |
| 30 | 5.09±0.34 | 3.26±0.39 | 4.46±0.27 | 3.55±0.08 |

**Table S8**: Electric energy consumption** (E, kWh m$^{-3}$) for the removal of *E. coli* in the experiments with continuous and intermittent current application in 10 mM phosphate buffer (PB) (anodic current, I$_{AN}$=200 mA) and tap water (TW) (I$_{AN}$=50 mA and 100 mA).

| Experiment | Continuous | 105 s ON 105 s OFF | 75 s ON 30 s OFF | 52.5 s ON 52.5 s OFF |
|---|---|---|---|---|
| PB, I$_{AN}$=200mA | 8.83±0.36 kWh m$^{-3}$, 4.9±0.2 log removal | 3.78±0.07 kWh m$^{-3}$, 3.1±0.2 log removal | 4.82±0.14 kWh m$^{-3}$, 4.2±0.2 log removal | 3.70±0.19 kWh m$^{-3}$, 3.2±0.4 log removal |
| TW, I$_{AN}$=50mA | - | - | 1.38±0.14 kWh m$^{-3}$, 4.3±0.2 log removal | - |
| TW, I$_{AN}$=100mA, | 5.70±0.24 kWh m$^{-3}$, 5.6±0.1 log removal | - | 4.55±0.01 kWh m$^{-3}$, 3.9±1.3 log removal | - |

**$**E = \frac{U*I}{q}$, where U-total cell potential (V), I-applied current (A), q-electrolyte flowrate (L h$^{-1}$)

**Table S9.** Characteristics of the employed tap water. $T_{AC}$ – total alkalinity, TOC- total organic carbon.

| Conductivity (µS cm$^{-1}$) | $T_{AC}$ (mg L$^{-1}$) | Cl$^-$ (mg L$^{-1}$) | $SO_4^{2-}$ (mg L$^{-1}$) | Na$^+$ (mg L$^{-1}$) | Mg$^{2+}$ (mg L$^{-1}$) | Ca$^{2+}$ (mg L$^{-1}$) | TOC (mg L$^{-1}$) |
|---|---|---|---|---|---|---|---|
| 450 | 130.1 | 21.4 | 13.4 | 13.8 | 9.2 | 51.2 | 2.1 |

**Table S10.** Recorded anode potentials ($E_{AN}$, V/SHE) and total cell potentials ($E_{TOT}$,V) at different applied anodic currents ($I_{AN}$, mA), in continuous and intermittent current mode using 10 mM phosphate Buffer (PB) and tap water (TW). Ohmic-drop in TW was calculated 1.7 and 3.4 V for 50 and 100 mA respectively. The present values have not been corrected for the ohmic-drop and represent outputs recorded by the potentiostat.

|  | PB, $I_{AN}$=200 mA |  | TW, $I_{AN}$=100 mA |  | TW, $I_{AN}$=50 mA |  |
|---|---|---|---|---|---|---|
|  | $E_{AN}$, V/SHE | $E_{TOT}$, V | $E_{AN}$, V/SHE | $E_{TOT}$, V | $E_{AN}$, V/SHE | $E_{TOT}$, V |
| **Continuous current** | 5.4±0.9 | 15.9±1.8 | 5.8±1.2 | 21±1.7 | - | - |
| **105 s ON - 105 s OFF** | 6.3±0.2 | 14.4±0.2 | - | - | - | - |
| **75 s ON - 30 s OFF** | 5.6±0.1 | 13.1±0.3 | 8.9±1.7 | 22.5±2.8 | 4.9±0.6 | 13.2±1.2 |
| **52,5 s ON - 52,5 s OFF** | 6.6±0.4 | 14.1±0.6 | - | - | - | - |

**Table S11.** Recorded anode potentials ($E_{AN}$, V/SHE) and total cell potentials ($E_{TOT}$,V) at different applied anodic currents ($I_{AN}$, mA) for various reactor configurations and flow directions, using 10 mM phosphate buffer (pH 7). Ohmic-drop in 10 mM phosphate buffer was determined to be 0.75, 1.5 and 3 V for 50, 100 and 200 mA, respectively, using electrochemical impedance spectroscopy (EIS). The present values have not been corrected for the ohmic-drop and represent outputs recorded by the potentiostat.

|  | $I_{AN}$=50 mA |  | $I_{AN}$=100mA |  | $I_{AN}$=200 mA |  |
|---|---|---|---|---|---|---|
|  | $E_{AN}$, V/SHE | $E_{TOT}$, V | $E_{AN}$, V/SHE | $E_{TOT}$, V | $E_{AN}$, V/SHE | $E_{TOT}$, V |
| **NRGO (A) - NRGO (C)** | 2.7±0.1 | 6.7±0.5 | 3.7±04 | 10.1±1.1 | 5.4±0.9 | 15.9±1.8 |
| **BRGO (A) - NRGO (C)** | 3.2±0.4 | 6.6±0.7 | 4.5±1.1 | 10.3±1.9 | 6.3±1.3 | 14.4±1.3 |
| **NRGO (C) - NRGO (A)** | 2.6±0.1 | 5.8±0.3 | 3.5±0.2 | 7.4±0.7 | 4.9±0.3 | 9.5±1.1 |
| **NRGO (C) - BRGO (A)** | 3.1±0.3 | 6.0±0.4 | 4.2±0.4 | 7.9±0.7 | 5.9±1.0 | 11.0±1.0 |